\documentclass[a4paper,11pt]{article}

\usepackage{amsmath}
\usepackage{amscd}
\usepackage{amsfonts}
\usepackage[]{graphicx}
\newcommand{\ee}{\end{eqnarray}}
\newcommand{\be}{\begin{eqnarray}}

\newcommand{\arcth}{\operatorname{arcth}}

\begin{document}

\begin{titlepage}
\bigskip
\def\thefootnote{\fnsymbol{footnote}}

\begin{center}
{\Large {\bf A and B branes from N=2 superspace
 } }
\end{center}

\vspace{1cm}

\begin{center}
{\large
{\text{{\bf Alexander Sevrin$^{1\, \ast}$, Wieland Staessens$^{1\, \ast\ast}$ and Alexander Wijns$^{2\, \ast \ast \ast}$}}}}
\end{center}

\begin{center}
\vspace{1em}
{\em  { $^{1}$ Theoretische Natuurkunde, Vrije Universiteit Brussel \\
\& The International Solvay Institutes,\\ 
Pleinlaan 2, B-1050 Brussels, Belgium.}}

\vspace{12pt}
 {\em  { $^{2}$ Department of Mathematics,\\
Science Institute, University of Iceland,\\ 
Dunhaga 2, 107 Reykjavik, Iceland.}} 
\end{center}

\vspace{2cm}

\noindent
\begin{center} {\bf Abstract} \end{center}
We present a manifestly supersymmetric description of A and B
branes on K\"ahler manifolds using a completely local $N=2$ superspace formulation
of the world-sheet nonlinear $\sigma$-model in the presence of a
boundary. In particular, we show that an $N=2$ superspace
description of type A boundaries is possible. This leads to a concrete realization of the still poorly understood coisotropic A branes. We also discuss briefly how the superspace description of a B brane provides an efficient way to compute higher loop $\beta$-functions. In particular, we sketch how one obtains the fourth order derivative correction to the Born-Infeld action by using a $\beta$-function method.

\vspace{.4cm} 

 \noindent Contribution to the Proceedings of the Third Workshop of the RTN project `Constituents, Fundamental Forces and Symmetries of the Universe', Valencia, October 1 - 5, 2007.

\vspace{3cm}

{\footnotesize $^\ast$ E-mail: Alexandre.Sevrin@vub.ac.be}

{\footnotesize $^{\ast\ast}$ E-mail: Wieland.Staessens@vub.ac.be}

{\footnotesize $^{\ast\ast\ast}$ E-mail: awijns@raunvis.hi.is}

\end{titlepage}

\section{Introduction and motivation}

In the context of compactifications of type II string theory in the presence of NS-NS fluxes, it is useful to study two-dimensional non-linear $\sigma$-models with $N=(2,2)$ supersymmetry. The extended supersymmetry on the world-sheet constrains the geometry of target space and an off-shell formulation of the $\sigma$-model helps characterizing the allowed geometries. It was recently shown that in an $N=(2,2)$ superspace formulation only chiral, twisted chiral and semi-chiral superfields are needed in order to parametrize the most general allowed target space \cite{Lindstrom:2005zr}. The most general allowed geometry giving rise to $N=(2,2)$ world-sheet supersymmetry is called bihermitian \cite{Gates:1984nk} or -- in the language of generalized complex geometry -- generalized K\"ahler \cite{Gualtieri:2003dx}. 

In the presence of D-branes one has to allow for $\sigma$-models with boundaries. A boundary only preserves a combination of the left- and right-handed supersymmetries, so that the relevant $\sigma$-model has what we will call $N=2$ supersymmetry. Although supersymmetric $\sigma$-models with boundaries have been the subject of much investigation, a completely local and off-shell description of the full $N=2$ supersymmetry is not known. Here, we present a new approach to an $N=2$ ``boundary superspace'' formulation of open string $\sigma$-models \cite{Koerber:2003ef,Sevrin:2007yn}. As a first step, we only allow for geometries which are parameterized by chiral or twisted chiral superfields exclusively. As we will see, this leads to a local, off-shell world-sheet description of both type A and type B branes on K\"ahler manifolds. Especially for A branes of coisotropic type this is the first time that such a description appears. The results presented in this text are currently being extended to manifolds which are locally parameterized by a combination of both chiral and twisted chiral fields \cite{wip}. This describes supersymmetric branes on a restricted class of generalized K\"ahler manifolds. 

Another setting in which superspace formulations are useful is for higher loop calculations in supersymmetric quantum field theories. The low energy effective field theory equations of motion of string theories can be obtained by calculating the world-sheet $\beta$-functions and demanding that they vanish. For example, the $\beta$-function of an $N=(2,2)$ superspace formulation of closed strings in a purely gravitational background was computed up to fourth loop order, leading to the $R^4$ correction to type II supergravity \cite{Grisaru:1986px}. Inspired by this success and using the proposed boundary superspace, an $N=2$ superspace calculation of the $\beta$-function was performed to obtain derivative corrections to D-brane low energy effective actions \cite{Nevens:2006ht}. Below, we briefly discuss the results of a three-loop $\beta$-function calculation, leading to the complete four-derivative correction to the Born-Infeld action.

\section{Boundary superspace}

To establish some notation and to explain the method, we first discuss how to go from $N=(1,1)$ to $N=1$ superspace \cite{Koerber:2003ef,Lindstrom:2002mc}. Consider the standard $N=(1,1)$ superspace $\sigma$-model action for a set of scalar fields $X^a$ parameterizing a target space with metric $g_{ab}$ and torsion proportional to $H=dB$, locally given in terms of a 2-form $B$, which is called the B-field,
 \be
 {\cal S}=2\int d^2 \sigma \, d \theta^+ d\theta^- \,D_+X^aD_-X^b\left(g_{ab}+b_{ab} \right).\label{an11}
 \ee
Here, $d^2\sigma = d\tau d\sigma$ is the integration measure over the (bosonic part of the) world-sheet, $\theta^\pm$ are Grassmann coordinates and $D_\pm$ are supercovariant derivatives. For our conventions, see \cite{Sevrin:2007yn}. This action is manifestly $N=(1,1)$ supersymmetric in the absence of boundaries. In the presence of a boundary, this is however no longer the case, since the boundary only preserves a linear combination of the supercovariant derivatives. Let us choose the boundary at $\sigma = 0$ in such a way that the supersymmetry corresponding to the combination $D = D_+ + D_-$ is preserved, while the one corresponding to $D' = D_+ - D_-$ is broken. These new supercovariant derivatives satisfy
 \be
 D^2=D'{}^2=- \frac{i}{2} \partial _ \tau ,\qquad \{D,D'\}=-i\, \partial _ \sigma.
 \ee
From these relations it follows that 
$-D\,D'=2 D_+\,D_- + i/2 \,\partial _ \sigma$, which shows that the action
 \be
 {\cal S}=-\int d^2 \sigma \, d \theta
\,D'\left(D_+X^aD_-X^b\left(g_{ab}+b_{ab}\right)\right),\label{an1}
 \ee
 differs from (\ref{an11}) only by a boundary term, while being manifestly invariant under the $N=1$ supersymmetry corresponding to $D$.

To make the boundary term in the variation of (\ref{an1}) disappear  one can either impose Dirichlet boundary conditions, $\delta X^a = 0$, or Neumann boundary conditions, $D'X^a = B^a{}_b DX^b$. To describe branes of intermediate dimensions one introduces projection operators,
 \be
 {\cal P}_\pm^a{}_b\equiv \frac{1}{2}\left( \delta ^a_b\pm R^a{}_b\right),
 \ee 
where $R^a{}_c\,R^c{}_b= \delta ^a_b$. Neumann and Dirichlet conditions are chosen along the eigenvectors of $ {\cal P}_+$ and $ {\cal P}_-$ respectively. The full set of conditions thus becomes (in matrix notation),
 \be
 {\cal P}_- \delta X = 0, \quad {\cal P}_+ D'X = {\cal P}_+ B {\cal P}_+ DX. \label{dn}
 \ee
For the last of these two equation, we assumed that $R_{ab} = R_{ba}$. For some subtleties involving this last issue, see \cite{Sevrin:2007yn}. Mixed boundary conditions of this type are only consistent if ${\cal P}_+$ is integrable.

The actions defined up till now can be written down for any Riemannian manifold (allowing for an almost product structure $R$ in case there is a boundary) as target space. Starting from these actions one can investigate for which target space geometries the $\sigma$-model is invariant under additional supersymmetry transformations \cite{Koerber:2003ef,Sevrin:2007yn}. We will not follow this route here. Instead we try to construct $N=(2,2)$ and $N=2$ supersymmetric actions directly in extended superspace. Adding two more Grassmann coordinates $\hat\theta^\pm$ and corresponding supercovariant derivatives $\hat D_\pm$ to the setup, the most general $N=(2,2)$ superspace action we can write down is
 \be
 {\cal S}=\int\,d^2 \sigma \,d\theta^+ \,d\theta^- \,d \hat \theta^+ \,d \hat \theta^- \, V(X), \label{an22}
 \ee
where $V$ is a scalar and dimensionless potential. As such, this action contains no dynamics. This is overcome by imposing constraints on the scalar fields. The most general linear constraints lead to the introduction of chiral, twisted chiral and semi-chiral superfields. Here, we focus on the first two classes. When the $\sigma$-model is defined in terms of chiral or twisted chiral fields exclusively, one easily shows that the resulting model has a target space which is K\"ahler with K\"ahler potential $V$ (and without torsion).

To describe D-branes on K\"ahler manifolds, we try to preserve half of the supersymmetry of this model in the presence of a boundary. To this end, we again introduce operators $D$, $D'$, $\hat D$ and $\hat D'$, where
 \be
 \hat D = D_+ \pm D_- \, , \quad \hat D' = D_+ \mp D_-\, . \label{dhat}
 \ee
We will take $D$ and $\hat D$ to correspond to preserved supersymmetries, while the other combinations are broken. The upper choice of signs in (\ref{dhat}) corresponds to what is called a B-type boundary and the lower choice corresponds to an A-type boundary. By the same kind of reasoning that led us to the $N=1$ action, we find that the action
 \be
 {\cal S}=\int d^2 \sigma\, d \theta d \hat \theta\, D' \hat D'\, V(X, \bar X)+
 i\,\int d \tau \,d \theta d \hat \theta \,W( X, \bar X), \label{an2}
 \ee
has manifest $N=2$ supersymmetry and differs from (\ref{an22}) only by a boundary term. Note that we were able to add a term with a boundary potential $W$, which will be crucial in what follows. It is quite easy to see that boundary conditions resulting from chiral fields in the presence of an A-type (B-type) boundary are completely equivalent to those described by twisted chiral fields in the presence of a B-type (A-type) boundary. Consequently, we can focus on B-type boundaries from now on without loss of generality. With the type of boundary fixed, the $\sigma$-model with only chiral fields will lead to a description of B branes, while twisted chiral fields will lead us to A branes. In this setting, mirror symmetry is nothing but an exchange of a chiral and twisted chiral fields, which indeed exchanges A and B branes.

\section{A branes}

Using the supercovariant derivatives appropriate for a B-type boundary, the constraint equations for twisted chiral superfields become (note that we use complex coordinates on target space from now on),
 \be
 \hat D X^ \mu = i D'X^ \mu ,\quad
 \hat D' X^ \mu = i DX^ \mu,\label{btcf}
 \ee
and the complex conjugate equations. Note that here $X^\mu$ and $D'X^\mu$ should be interpreted as two independent (and for the moment unconstrained) $N=2$ superfields. More precisely, they are different components of the $N=(2,2)$ superfield $X^\mu$ when expanded in terms of $\theta'$.\footnote{By a slight abuse of notation, we used the same symbol for the $N=(2,2)$ superfield $X^\mu$ and its first $N=2$ component $X^\mu \vert_{\theta' = \hat\theta' = 0}$.} To arrange for the boundary term in the variation of the action to vanish, we again impose Dirichlet boundary conditions of the form,
 \be
 \delta X^ \mu = R^ \mu {}_{ \bar \nu }\, \delta X^{ \bar \nu }+
 R^ \mu {}_{  \nu } \,\delta X^{  \nu }. \label{di1}
 \ee
These conditions imply similar relations for $\hat D X^\mu$. Consistency of these equations with the constraint equations (\ref{btcf}) imply the existence of two new projection operators,
 \be
 \pi _+^ \mu {}_ \nu \equiv R^ \mu {}_ \nu , \quad
 \pi _-^ \mu {}_ \nu \equiv R^ \mu {}_{ \bar \rho }R^{ \bar \rho }{}_ \nu.
 \ee
In terms of these operators, the Neumann boundary conditions can be written as
 \be
 \left( \pi _+ {\cal P}_+D'X\right)^ \mu= R^ \mu {}_\nu D' X^ \nu , \quad
 \left( \pi _- {\cal P}_+D'X\right)^ \mu= 0.\label{lkj2}
 \ee
Comparing this to the $N=1$ Neumann conditions (\ref{dn}), we see that in the $\pi_+$ directions there is a non-degenerate $U(1)$ bundle, while in the $\pi_-$ directions there can only be a flat $U(1)$ connection.\footnote{The restricted class of models we discuss here have no torsion. Yet there can be a locally exact B-field, $B = F = dA$, with $A$ some $U(1)$ connection.}

When $\pi_- = 1$, {\em i.e.} $R^ \mu {}_ \nu = 0$, the condition (\ref{di1}) simplifies and it is easy to see that each Dirichlet condition implies a Neumann condition. The boundary term in the variation of the action vanishes if
 \be
 \left(V+i\,W\right)_ \mu R^ \mu{}_{ \bar \nu }=
 \left(V-i\,W\right)_{ \bar \nu } ,
 \ee
which implies that $R_{ \mu \nu }= R_{ \nu \mu }$. This in turn implies that the pull-back of the K\"ahler form to the brane vanishes. It follows that the brane wraps an isotropic submanifold of the target space (seen as a symplectic manifold) of maximal dimension, {\em i.e.} a Lagrangian submanifold.

When on the other hand, $\pi_+ = 1$, the Dirichlet condition (\ref{di1}) becomes trivial and the brane is space filling. From (\ref{lkj2}) it follows that there should be a non-degenerate $U(1)$ field strength F such that the Neumann conditions are schematically of the form $D'X = F DX$. Because the $X^\mu$ are twisted chiral fields this implies that 
 \be
 \hat D X^a = K^a{}_b D X^b, \label{bc}
 \ee
where $K = -\omega^{-1} F$, with $\omega$ the K\"ahler form. Equation (\ref{bc}) has two important implications. First of all it is only consistent if $K$ is an extra complex structure independent from the complex structure $J$ associated with $\omega$. Secondly, it constitutes a constraint equation for the $X^\mu$. It is thus important to make sure that these constraints are obeyed along the deformation when varying the action. This is most easily done in this case by defining the $X^\mu$ in terms of unconstrained fields in such a way that (\ref{bc}) is automatically satisfied. In this way it can be shown that the boundary term vanishes if the $U(1)$ field strength is locally given by $F=dA$, were the vector potential depends on both potentials $V$ and $W$, and the complex structures 
 \be
 A_a = - \frac 1 2 V_c (JK)^c{}_a +
 \frac 1 2 W_c K^c {}_a + \partial _a f,\label{eee5}
 \ee
defined up to some gradient of a real function $f$. From the anti-symmetry of $F$, we also find that both $F$ and $\omega$ are (2,0) + (0,2) forms with respect to the complex structure $K$. Since they are non-degenerate it follows that the target space is $4l$-dimensional, $l \in \mathbb{N}$. This type of brane is called a maximally coisotropic brane.

To treat the generic case where both $\pi_+$ and $\pi_-$ are nonzero, we assume that they are both integrable and diagonalizable. A new condition which follows from imposing the vanishing of the boundary variation is that $\omega$ has a block diagonal structure with respect to the splitting of the target space with respect to $\pi_\pm$. Apart from that, the generic situation is just a reasonably straightforward combination of the two extreme cases. In the $\pi_-$ directions, there are again an equal number of Dirichlet and Neumann conditions, and the pullback of $\omega$ vanishes. In the $\pi_+$ directions the brane is again space filling and $4l$-dimensional. This implies that a generic A brane on a K\"ahler manifold is $(n +2l)$-dimensional, where the target space dimension is $d=2n$ and $l \in \mathbb{N}$. These branes are called coisotropic for the following reason. Denoting the space tangent to the submanifold wrapped by the brane (at a certain point) by $T$ and its symplectic orthogonal by $T^\bot$, the submanifold is called coisotropic if $T^\bot \subset T$. From the block diagonal structure of $\omega$ and the fact that it vanishes in the image of $\pi_-$, this is easily shown to be the case for A branes. More precisely, $T/T^\bot$ is the submanifold spanned by the $\pi_+$ eigenvectors. In accordance with the literature \cite{Lindstrom:2002jb,Kapustin:2001ij}, we showed that the non-degenerate $U(1)$ field strength $F$ and the complex structure $K = -\omega^{-1} F$ live on $T/T^\bot$, which is necessarily $4l$-dimensional. The Lagrangian and maximally coisotropic branes are clearly special cases of this more general structure.

\section{B branes}

Since an $N=2$ boundary superspace description of B branes was already constructed before \cite{Koerber:2003ef} and B branes are much better understood in the literature in general, we will only briefly sketch the general formalism in this case and focus more on the application to loop calculations. This time we need chiral superfields in the presence of a B-type boundary, 
 \be
 \hat D X^ \alpha = i DX^ \alpha ,\quad
 \hat D' X^ \alpha = i D'X^ \alpha ,\label{bcf}
 \ee
and their complex conjugates. Unlike in the twisted chiral case, these immediately imply that the $X^\alpha$ are not independent fields. Again we solve for the constraints by expressing the $X^\alpha$ in terms of unconstrained superfields $\Lambda^\alpha$, $X^\alpha = (\hat D - i D) \Lambda^\alpha$. Imposing a Dirichlet-like condition 
 \be
 \delta \Lambda ^ \alpha = R^ \alpha {}_ \beta\, \delta \Lambda ^ \beta +
 R^ \alpha {}_{ \bar \beta } \,\delta \Lambda ^{ \bar \beta },
 \ee
implies an analogous condition for $\delta X^\alpha$ by integrability of ${\cal P}_+$. Consistency with the chirality conditions (\ref{bcf}) enforces $R^ \alpha {}_{ \bar \beta } = 0$, {\em i.e.} we get
$ \delta X^ \alpha =R^ \alpha {}_ \beta \,\delta X^ \beta.$ 
This shows that the Dirichlet projection commutes with the complex structure, so that the submanifold wrapped by a B brane is a holomorphic submanifold of the K\"ahler manifold. Working out the boundary term in the variation of (\ref{an2}) leads to boundary conditions of the form (\ref{dn}), albeit with holomorphic projections. The $U(1)$ connection needs to be that of a holomorphic vector bundle
 \be
 F_{ \alpha \beta }= F_{ \bar a \bar \beta }=0 \label{holo}
 \ee
and is related to the boundary potential $W$ by
 \be
 F_{ \alpha \bar \beta }= - F_{ \bar \beta \alpha }=-i\,W_{ \alpha \bar \beta }. \label{FW}
 \ee

The above characterizations of A and B branes follow solely from the $N=2$ supersymmetry on the world-sheet. In this sense they correspond to A and B branes in topological string theory. To ensure that they are consistent objects in the full string theory, we need to impose conformal invariance as well. As is well known, this is done by imposing the vanishing of the world-sheet $\beta$-functions for a string moving through a certain background, which results in equations of motion for the background. With our $N=2$ boundary superspace model at hand, we can apply the $\beta$-function method to find higher order corrections to the D-brane low energy effective action, {\em i.e.} the Born-Infeld action. The computation is fairly straight forward (albeit still tedious) when we consider a space-filling B brane in flat space, for which the action is that of a set of free chiral superfields $X^\alpha$ and fermionic superfields $D'X^\alpha$, coupled at the boundary to a potential $W(X)$, which is treated as an interaction in perturbation theory. Order by order in the number of loops, the $\beta$-function(al) corresponding to the coupling of the chiral superfields to the boundary potential is set to zero. Since the $U(1)$ field strength is related to $W$ by (\ref{FW}), this results in equations of motion for $F_{\alpha \bar\beta}$. One of the strengths of the $\beta$-function approach is that the number of loops $l$ corresponds to the number of derivatives $n$ on the field strength according to the relation $n = 2(l-1)$, and that at every order in the number of derivatives, the result is all order in $\alpha'$. This means that at one loop order, we should find the Born-Infeld action. Indeed, we find,
 \be
 \beta (W) \propto g^{\alpha\bar\beta} (\arcth F)_{\alpha \bar \beta} = 0 \quad \mbox{at one loop.}
 \ee
This is the equation of motion corresponding to the Born-Infeld action for a holomorphic vector bundle connection. It is called the deformed stability condition, because it is a deformation of the Donaldson-Uhlenbeck-Yau stability condition $g^{\alpha\bar\beta} F_{\alpha \bar \beta} = 0$, which together with (\ref{holo}) solves the Yang-Mills equations of motion. At two loops, we find no contribution to the $\beta$-function, so that there are no two-derivative corrections to the Born-Infeld action, which is in agreement with the literature. At third loop order, we do find a contribution, such that the deformed stability condition up to four derivatives becomes \cite{Koerber:2004ze},
 \be
  g^{ \alpha \bar \beta } \left( \arcth\, F \right)_{ \alpha
   \bar \beta }
   + \frac{1}{96} S_{ab\alpha \bar
   \beta} S_{cd\gamma \bar \delta}\; h_+^{bc}h_+^{da}\left(
   h_+^{\alpha \bar \delta} h_+^{\gamma \bar \beta} -
   h_-^{\alpha \bar \delta} h_-^{\gamma \bar \beta} \right) = 0,
 \ee
where
$S_{abcd} = \partial_a \partial_b F_{cd} + 2 h_+^{ef} \partial_a
 F_{[c\vert e}\; \partial_b F_{\vert d]f}$ and $h^\pm_{ab} = \eta_{ab} \pm F_{ab}.$
This together with (\ref{holo}) solves the equations of motion for the action 
 \be
  {\cal S} &=& - \tau_9 \int d^{10}x\; \sqrt{-h_+}\bigg[ 1 +
   \frac{1}{96} \Big( \frac 1 2 h_+^{\mu\nu}h_+^{\rho\sigma}
   S_{\nu\rho}
   S_{\sigma\mu} \nonumber\\
   &&-
   h_+^{\rho_2\mu_1}h_+^{\mu_2\rho_1}h_+^{\sigma_2\nu_1}h_+^{\nu_2\sigma_1}
   S_{\mu_1\mu_2\nu_1\nu_2}S_{\rho_1\rho_2\sigma_1\sigma_2}\Big)\bigg],
 \ee
which agrees with the result in \cite{Wyllard:2000qe} derived from a boundary conformal field theory computation. For many more details on how to obtain this result, we invite the reader to consult \cite{Nevens:2006ht}.

\section{Further remarks and outlook}

We exhibited a completely local $N=2$ superspace description of A and B branes on K\"ahler manifolds. To achieve this, it was only necessary to obtain the necessary formalism for $\sigma$-models with chiral fields or twisted chiral fields exclusively. As was mentioned, the two possibilities are related by mirror symmetry. In \cite{Sevrin:2007yn} we obtained explicit duality transformations between certain A and B branes for situations when there is an isometry, by gauging the isometry and passing through a first order formalism. In particular, we found the explicit duality transformation between a space filling B brane on a K\"ahler manifold and the dual Lagrangian A brane on the dual K\"ahler manifold. A more interesting case which was achieved is a dualization along one of the two isometries of a maximal coisotropic brane on a four-dimensional hyperk\"ahler manifold, which leads to a 3-brane on a generalized K\"ahler manifold, a system that does not fit in two classes discussed here. To describe this system it is necessary to consider $\sigma$-models with both chiral and twisted chiral fields. This will lead to a description of D-branes on a certain class of generalized K\"ahler manifolds and is work in progress \cite{wip}. Eventually, to obtain a complete description of branes resulting from $N=2$ boundary conditions, semi-chiral fields will have to be considered as well.

A next step, which connects to the previous section, is obtaining the stability conditions for more general branes (coisotropic branes, branes wrapping generalized complex submanifolds of generalized K\"ahler manifolds) by demanding one (and if feasible, higher) loop conformal invariance of the corresponding $\sigma$-models.

\section*{Acknowledgments}
All authors are supported in part by the European Commission FP6 RTN
programme MRTN-CT-2004-005104. AS and WS
are supported in part by the Belgian Federal Science Policy Office through
the Interuniversity Attraction Pole P6/11,  and in part by the
``FWO-Vlaanderen'' through project G.0428.06. AW is supported in part by grant 070034022
from the Icelandic Research Fund.

\end{document}